\documentclass[journal=ancac3]{achemso}
\usepackage[version=3]{mhchem}

\author{Sergey Makarov}
\affiliation[ITMO University]
{Laboratory of Nanophotonics and Metamaterials, ITMO University, St.~Petersburg, Russia}
\email{s.makarov@metalab.ifmo.ru}

\author{Sergey Kudryashov}
\affiliation[Lebedev Physical Institute]
{Lebedev Physical Institute, Moscow, Russia}

\author{Ivan Mukhin}
\affiliation[ITMO University]
{Laboratory of Nanophotonics and Metamaterials, ITMO University, St.~Petersburg, Russia}
\alsoaffiliation[St. Petersburg Academic University]
{Laboratory of Renewable Energy Sources, St. Petersburg Academic University, St.~Petersburg, Russia}

\author{Alexey Mozharov}
\affiliation[St. Petersburg Academic University]
{Laboratory of Renewable Energy Sources, St. Petersburg Academic University, St.~Petersburg, Russia}

\author{Valentin Milichko}
\affiliation[ITMO University]
{Laboratory of Nanophotonics and Metamaterials, ITMO University, St.~Petersburg, Russia}

\author{Alexander Krasnok}
\affiliation[ITMO University]
{Laboratory of Nanophotonics and Metamaterials, ITMO University, St.~Petersburg, Russia}
\email{krasnokfiz@mail.ru}

\author{Pavel Belov}
\affiliation[ITMO University]
{Laboratory of Nanophotonics and Metamaterials, ITMO University, St.~Petersburg, Russia}
\email{belov@phoi.ifmo.ru}

\title[An \textsf{achemso} demo] {Tuning of magnetic optical response in a dielectric nanoparticle by ultrafast photo-injection of dense electron-hole plasma}


\begin{document}

\begin{tocentry}
\includegraphics[width=1.6\columnwidth]{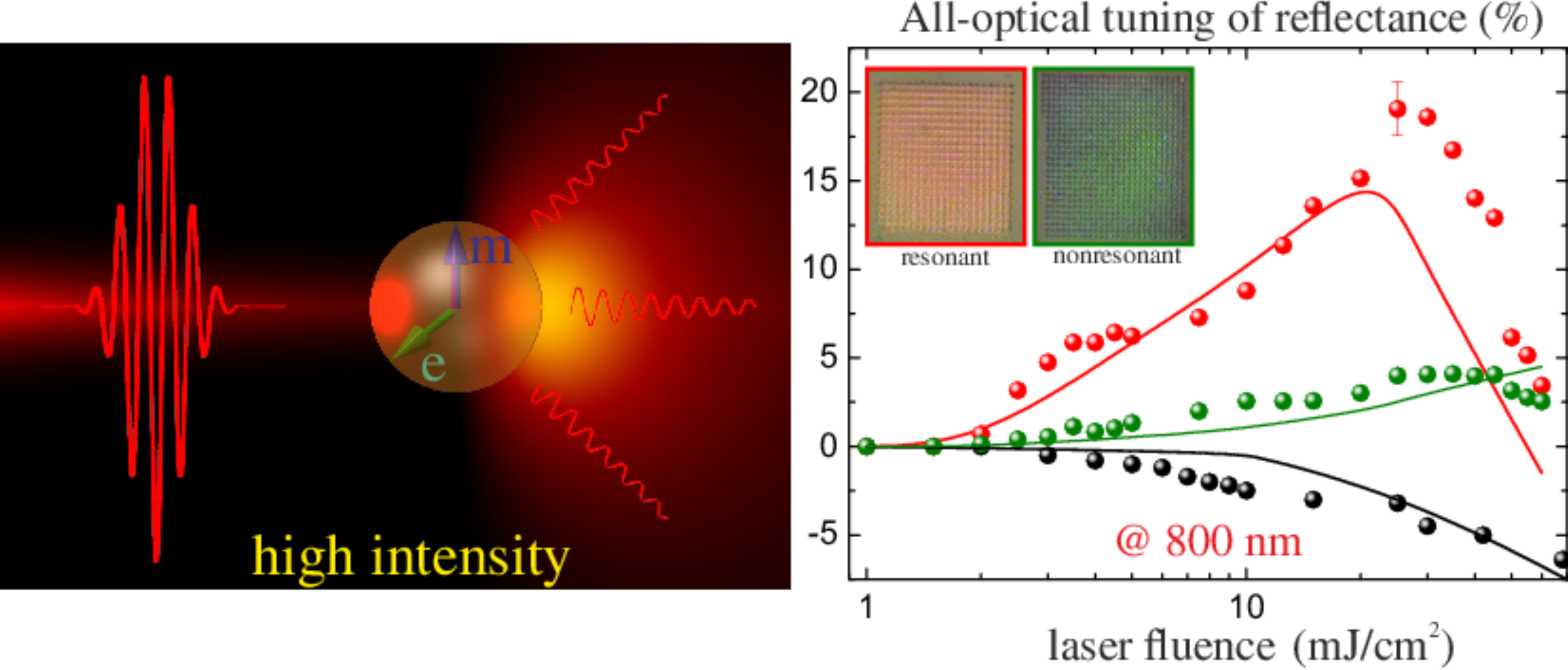}
\end{tocentry}

\begin{abstract}
We propose a novel approach for efficient tuning of optical properties of a high refractive index subwavelength nanoparticle with a magnetic Mie-type resonance by means of femtosecond laser irradiation. This concept is based on ultrafast photo-injection of dense ($>10^{20}$ cm$^{-3}$) electron-hole plasma within such nanoparticle, drastically changing its transient dielectric permittivity. This allows to manipulate by both electric and magnetic nanoparticle responses, resulting in dramatic changes of its scattering diagram and scattering cross section. We experimentally demonstrate 20$\%$ tuning of reflectance of a single silicon nanoparticle by femtosecond laser pulses with wavelength in the vicinity of the magnetic dipole resonance. Such single-particle nanodevice enables to design fast and ultracompact optical switchers and modulators.
\end{abstract}

All-dielectric "magnetic light" nanophotonics based on nanoparticles of high refractive index materials allows manipulation of a magnetic component of light at nanoscale without high dissipative losses, inherent for metallic nanostructures~\cite{Cummer_08, Zhao09, evlyukhin2010, kuznetsov2012, Evlyukhin:NL:2012, Miroshnichenko:NL:2012, Brener_12,  krasnok2015towards}. This "magnetic light" concept has been implemented for nanoantennas~\cite{KrasnokOE}, photonic topological insulators~\cite{Slobozhanyuk2015}, broadband perfect reflectors~\cite{Krishnamurthy_13}, waveguides~\cite{Savelev2014_1}, cloacking~\cite{cloaking2015all}, harmonics generation~\cite{ShcherbakovNL2014}, wave-front engineering and dispersion control~\cite{Staude_15}.

Such magnetic optical response originates from circular displacement currents excited inside the nanoparticle by incident light. This opens the possibility of interference between magnetic and electric modes inside the dielectric nanoparticle at certain wavelength. One of the most remarkable effects based on this concept is formation of the so-called Huygens source, scattering forward the whole energy~\cite{kerker1983}, while for another wavelength range the nanoparticle can scatter incident light almost completely in backward direction~\cite{krasnok2011huygens, Lukyanchuk13}. Therefore, manipulation by both electric and magnetic resonances paves the way for efficient tuning of the dielectric nanoparticle scattering in the optical range. The spectral positions of the electric and magnetic dipole resonances depend on the particle shape and environment~\cite{evlyukhin2010, Brener_12, Evlyukhin:NL:2012, EvlyukhinSciRep2014, yang2014all, Staude_15}. Alternatively, these resonances can be tuned permanently via changes in dielectric permittivity, which was achieved by annealing of amorphous silicon nanoparticles~\cite{chichkov2014NatCom}.

However, modern optical technologies require fast, large, and reversible modulation of optical response of ultracompact functional nanostructures. For this purpose, different types of optical nonlinearities both in metallic~\cite{zayats2012nonlinear} and dielectric structures~\cite{LPR2015review} were utilized such as Kerr-type nonlinearities~\cite{zhou2010analytical, Noskov12, abb2014hotspot}, free carriers generation~\cite{mazurenko2003ultrafast, dani2009subpicosecond, large2010} and variation of their temperature~\cite{zayats2011fsnanorods}, as well as relatively slow thermal nonlinearity~\cite{notomi2005optical}. Since plasmonic structures have high inherent losses, while photonic crystals or graphene-based structures~\cite{graphene2014ultrafast} are much larger than the wavelength, it is advantageous to use low-loss and subwavelength high-index particles with electric and magnetic responses (the "magnetic light" concept). Moreover, nonlinear manipulation by scattering characteristics (power pattern and cross section) of a nanostructure via its magnetic response gives an additional efficient tool for ultrafast all-optical switching and routing.

\begin{figure}[!t]
\includegraphics[width=0.9\textwidth]{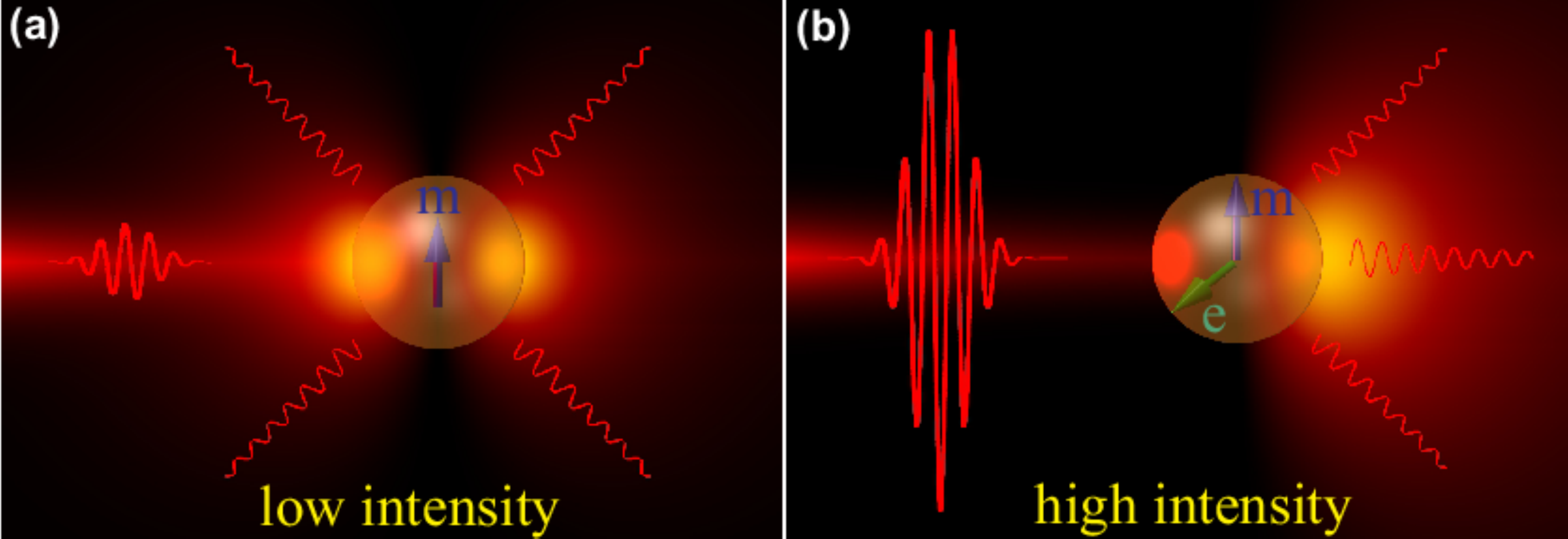}
\caption{A schematic illustration of scattering manipulation by a single weak (\textbf{a}) and intense (\textbf{b}) femtosecond laser pulse. Intense femtosecond laser pulse switch-on a Huygens source regime, when incident light is scattered in the forward direction.}\label{Fig1}
\end{figure}

In this work we propose a novel approach for manipulation by the electric and magnetic responses of a high-index dielectric nanoparticle, employing its ultrafast photoexcitation by femtosecond laser irradiation. Specifically, we demonstrate theoretically possibility of large tuning of scattering properties of the single nanoparticle (near the regime of the Huygens source as shown in Fig.~\ref{Fig1}) and achieve experimentally 20 $\%$ changes of reflection from silica surface with a silicon nanoparticle under femtosecond laser irradiation.

Femtosecond (fs) laser pulses are known to provide strong photo-induced electronic excitation of free carriers in diverse materials (owing to usually low electronic heat capacities), which is accompanied by sub-10 fs thermalization of free carriers and dramatic variation of material optical characteristics during the pumping fs-laser pulse (usually $\leq$ 100 fs). Such almost prompt, fs-laser induced optical tunability appears to be much broader for (semi)insulating materials with very minor initial carrier concentrations, extending in a sub-ablative regime in the visible and near-IR ranges.
In detail, such ultrafast transient modulation of optical dielectric permittivity in semiconductors and dielectrics is related to transient variation of free-carrier (electron-hole plasma, EHP) density $\rho_{\rm eh}$ through its basic intraband and interband contributions~\cite{Hirlimann83, Antonetti84, Downer90, Mazur95, Linde2000, Kudryashov2002}.
Simultaneously, ultrafast transient optical modulation is additionally enhanced due to a strong prompt EHP-driven isotropic renormalization of direct bandgap, resulting in drastic enhancement of interband transitions and corresponding red spectral shift of the optical dielectric permittivity~\cite{Mazur95, ionin2012ultrafast}. Remarkably, such strong optical modulation in semiconductors, requiring high EHP density $\rho_{\rm eh}>10^{20}$ cm$^{-3}$, is rapidly reversible on a picosecond time scale, e.g., down to $\sim$6--7 ps for c-Si owing to three-body Auger recombination~\cite{yoffa1980, Hirlimann83} or even down to $\sim$1~ps for \textit{a}-Si~\cite{fauchet1992properties}.

In this work a realistic pump-pulse averaged dependence of the dielectric permittivity for the photo-excited silicon versus incident fs-laser fluence at 800-nm laser wavelength was obtained through modeling and fitting of experimental data from our previous work~\cite{Kudryashov2002} (Fig.~\ref{Fig2}), which is well consistent with data from many other papers~\cite{Hirlimann83, Antonetti84, Downer90, Linde2000}. In particular, we analyze single-shot fs-laser pump self-reflectivity (\textit{R}) from an atomically smooth silicon surface at its \textit{s}- $(R_s(45^\circ))$ and \textit{p}- $(R_p(45^\circ))$ polarizations at the $45^\circ$-incidence angle and variable effective (absorbed) laser fluences $F_{\rm eff}=(1-R_{\rm s,p}(45^\circ,F))$$\cdot$\textit{F}, where \textit{F} is the incident fluence, using a model transient dielectric permittivity for photo-excited silicon. Commonly, such model dielectric permittivity, being a function of $\rho_{\rm eh}$, can be expressed as a sum of interband- and intraband-transition based terms~\cite{Linde2000, ionin2012ultrafast}:
\begin{figure}[!t]
\includegraphics[width=0.8\textwidth]{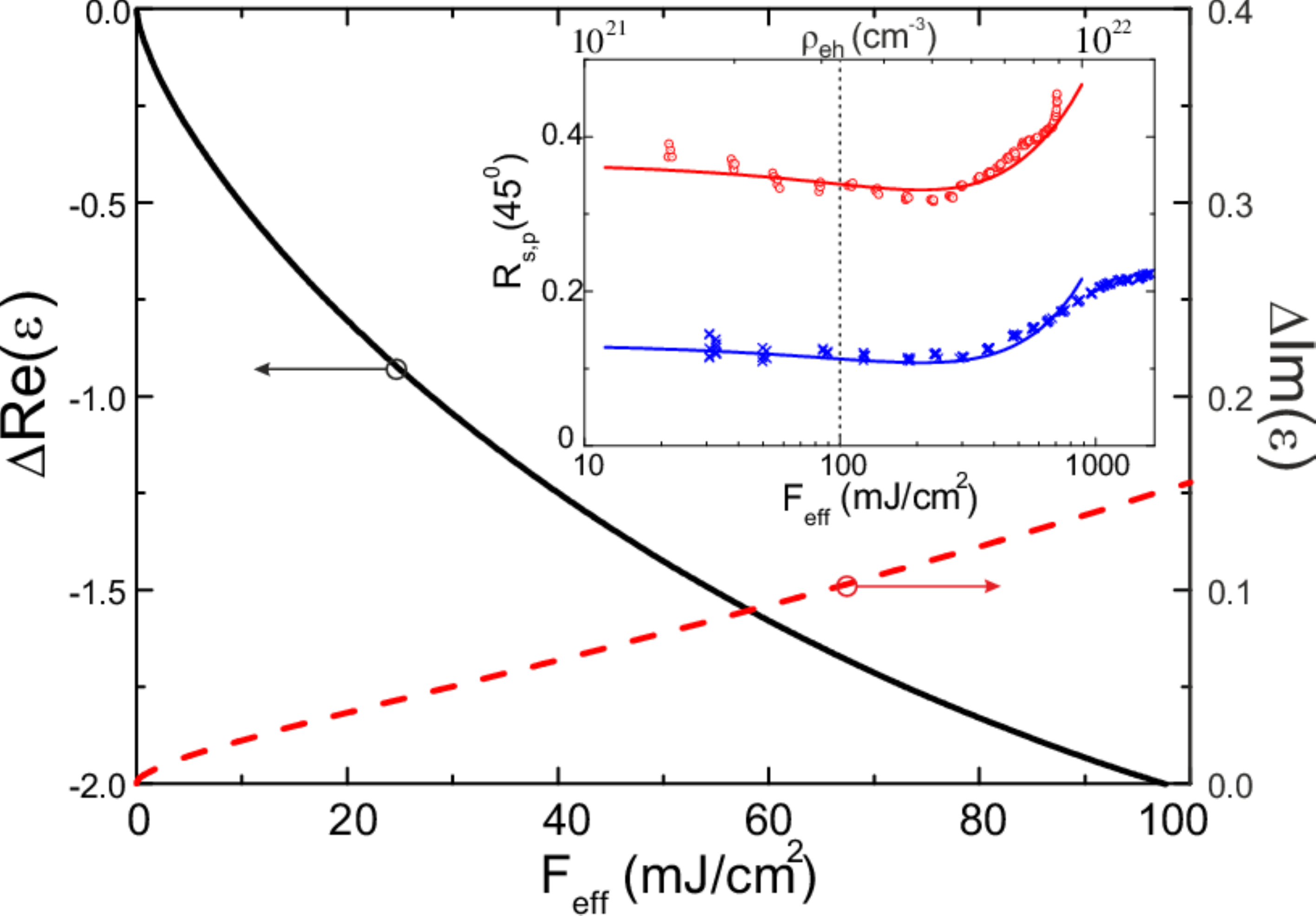}
\caption{Derived changes of real (black curve) and imaginary (red curve) components of optical dielectric permittivity versus $F_{\rm eff}$ at the 800-nm pump wavelength relatively their initial values. Inset: experimental pump self-reflectivity dependencies $R_{\rm s,p}(45^\circ,F_{\rm eff})$ (red and blue dots, respectively, adapted from~\cite{Kudryashov2002}) on effective laser fluence with their corresponding fitting red and blue model curves $R_{\rm s,p}(45^\circ,\rho_{\rm eh})$. The black dashed line shows a typical value of laser fluence, which is enough for strong changing of optical properties of bulk silicon or a silicon nanoparticle (see Fig. 4) at sub-damage conditions.}\label{Fig2}
\end{figure}
\begin{eqnarray}
\varepsilon(\omega,\rho_{\rm eh})=\varepsilon_{\rm IB}(\omega^*)\left(1-\frac{\rho_{\rm eh}}{\rho_{\rm bf}}\right)-\frac{\omega^2_{\rm pl}(\rho_{\rm eh})}{\omega^2+1/(\tau^2_e(\rho_{\rm eh}))}\left(1-\frac{i}{\omega\tau_e(\rho_{eh})}\right),
\end{eqnarray}
where the above mentioned $\rho_{\rm eh}$-dependent bandgap shrinkage effect on interband transitions is accounted by introducing effective photon frequency $\omega^{*}= \omega+\Theta\rho_{\rm eh}/\rho_{\rm bgr}$ with the factor $\Theta$, the characteristic renormalization EHP density $\rho_{\rm bgr}\approx1\times10^{22}$~cm$^{-3}$~\cite{ionin2012ultrafast}, being typically about 5$\%$
of the total valence electron density $\approx2\times$10$^{23}$~cm$^{-3}$ in Si) to provide the ultimate 50$\%$
electronic direct bandgap renormalization~\cite{Louie2004}, i.e., $\hbar\Theta\approx$ 1.7~eV of the effective minimal gap $\approx3.4$~eV in silicon~\cite{palik, Dargys_book}, while $\rho_{\rm bf}$ is the characteristic band capacity of the specific photo-excited regions of the first Brillouine zone in the k-space (e.g., $\rho_{\rm bf}(L)\approx4\times10^{21}$~cm$^{-3}$ for L-valleys and $\rho_{\rm bf}(X)\approx 4.5\times10^{22}$ cm$^{-3}$ for X-valleys in Si), affecting interband transitions via the band-filling effect~\cite{Mazur95, ionin2012ultrafast, Downer90, Linde2000}. The bulk EHP frequency $\omega_{\rm pl}$ is defined as
\begin{equation}\label{S4}
\omega_{\rm pl}^{2}(\rho_{\rm eh})=\frac{\rho_{\rm eh}e^2}{\varepsilon_0\varepsilon_{\rm hf}(\rho_{\rm eh})m^{*}_{\rm opt}(\rho_{\rm eh})},
\end{equation}
where the effective optical (e-h pair) mass $m^{*}_{\rm opt}\approx0.14$$m_{\rm e}$ in L-valleys or 0.19 in X-valleys~\cite{ionin2012ultrafast, Antonetti84, Linde2000, Dargys_book} is a $\rho_{\rm eh}$-dependent quantity, varying versus transient band filling due to the band dispersion and versus bandgap renormalization~\cite{Cardona2005}. The high-frequency electronic dielectric constant $\varepsilon_{\rm hf}$ was modeled in the form $\varepsilon_{\rm hf}(\rho_{\rm eh})=1+\varepsilon_{\rm hf}(0)\times\exp(-\rho_{\rm eh}/\rho_{\rm scr})$, where the screening density $\rho_{\rm scr}\approx1\times10^{21}$~cm$^{-3}$ was chosen to provide $\varepsilon_{\rm hf}\rightarrow1$ in dense EHP. The electronic damping time $\tau_e$ in the regime of dense EHP at the probe frequency $\omega_{\rm pr}$ was taken, similarly to metals, in the random phase approximation as proportional to the inverse bulk EHP frequency $\omega_{\rm pl}^{-1}$~\cite{Lagendijk1995}
\begin{equation}\label{S5}
\tau_e=\left(\frac{128E_{F}^2}{\pi^2\sqrt{3}\omega_{\rm pl}}\right)\frac{1+\exp\left[\frac{\hbar\omega}{k_{B}T_{e}}\right]}{(\pi k_{B}T_{e})^2+(\hbar\omega)^2},
\end{equation}
where $E_{\rm F}\approx$ 1 -- 2~eV is the effective Fermi-level quasi-energy for electrons and holes at $\rho_{\rm eh}<1\times10^{22}$~cm$^{-3}$, $\hbar$ and $k_B$ are the reduced Planck and Boltzmann constants, respectively, and $T_{\rm e}$ is the unified EHP temperature, being a weak function of $\rho_{\rm eh}$~\cite{yoffa1980}. Here, the latter relationship was evaluated for $\hbar\omega>k_{\rm B}T_{\rm e}$ in the form $\tau_{\rm e}(\rho_{\rm eh})\approx3\times10^{2}/(\omega_{\rm pl}(\rho_{\rm eh}))$, accounting multiple carrier scattering paths for the three top valence sub-bands, and multiple X-valleys in the lowest conduction band of silicon.

The resulting $\rho_{\rm eh}$-dependent oblique-incidence pump reflectivities $R_{\rm s,p}(45^\circ,\rho_{\rm eh})$ model, calculated using common Fresnel formulae, fit well the extracted experimental reflectivity dependences $R_{\rm s,p}(45^\circ,F_{\rm eff})$ in Fig.~2 with the characteristic initial dip and the following rise. Such reasonable fitting in Fig.~\ref{Fig2} provides an important relationship between magnitudes $\rho_{\rm eh}$ and $F_{\rm eff}$ in the region, covering the reflectivity dip and rise, which were used to plot the derived optical dielectric permittivity versus $F_{\rm eff}$ (Fig.~\ref{Fig2}).

Such large changes of optical dielectric permittivity properties in silicon at fluences below its melting (\textit{F}$_{\rm eff}$~$\approx$~0.17~J/cm$^2$~\cite{ionin2013Si}) and ablation thresholds (for spallation under these experimental conditions, \textit{F}$_{\rm eff}$~$\approx$~0.3~J/cm$^2$~\cite{ionin2013Si}) can significantly alter optical response of a silicon nanoparticle, supporting a magnetic Mie-type resonance. Using the extracted dielectric permittivity values of photoexcited silicon, we study such optical tuning of scattering properties of a silicon nanoparticle in vacuum near its magnetic resonance by means of full-wave numerical simulations carried out in CST Microwave Studio.

\begin{figure}[!t]
\includegraphics[width=0.8\textwidth]{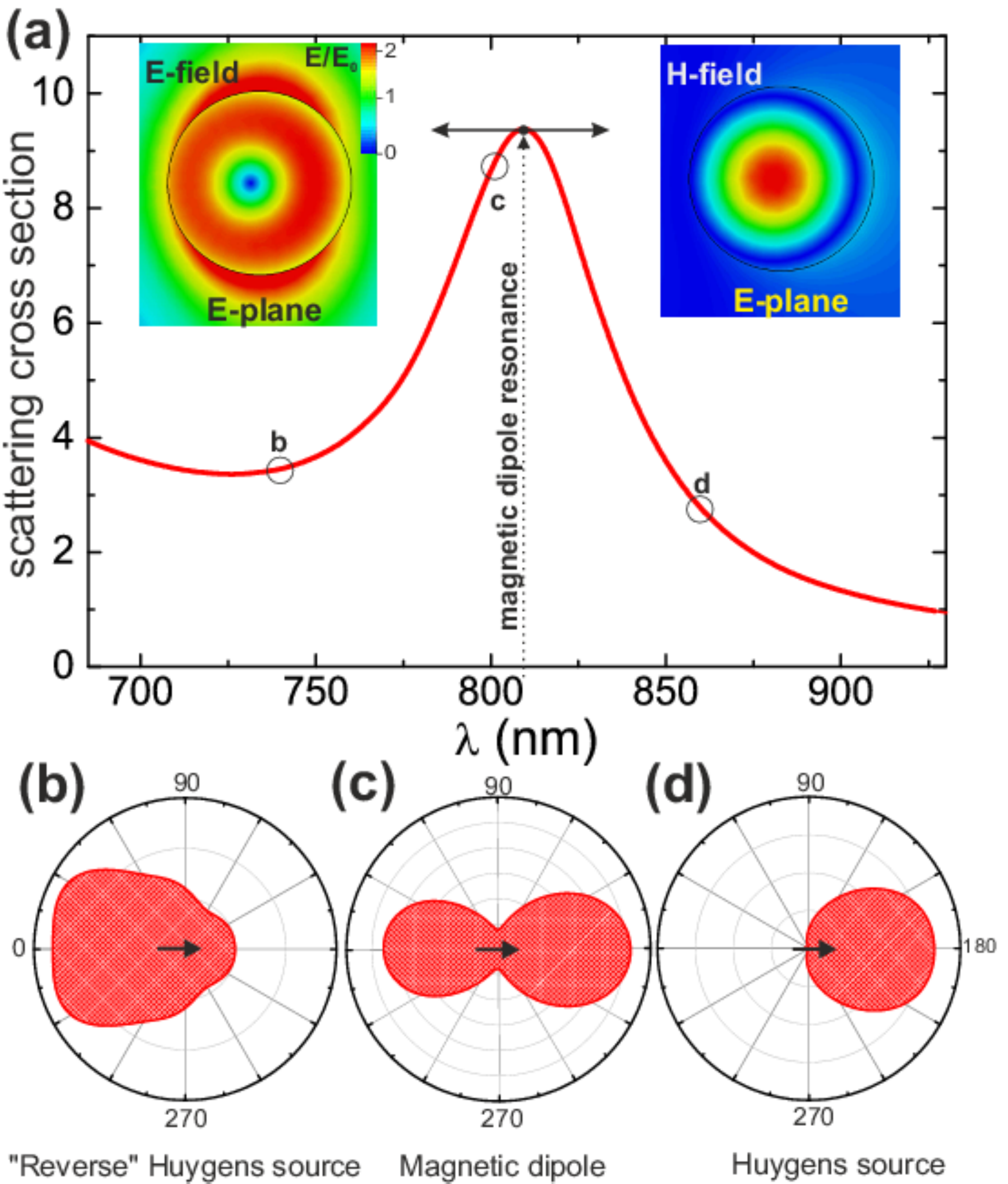}
\caption{Numerically calculated scattering spectrum of a silicon sphere of a diameter \textit{D} = 210~nm (\textbf{a}). Insets: numerically calculated electric and magnetic fields distributions inside and near the silicon sphere. Scattering diagrams of the silicon sphere at $\lambda$ = 740~nm (\textbf{b}), 800~nm (\textbf{c}) and 860~nm (\textbf{d}). The black arrows in (\textbf{b}--\textbf{d}) indicate the direction of light incidence.}\label{Fig3}
\end{figure}

We numerically analyzed optical properties of a spherical (the diameter \textit{D} $\equiv$ 2\textit{r} = 210~nm) silicon particle with its dilectric permittivity, depending on laser fluence as shown in Fig.~\ref{Fig1}b. The chosen nanoparticle diameter corresponds to excitation of a magnetic dipole Mie-type resonance in the vicinity of the femtosecond laser wavelength $\lambda$ $\approx$ 800~nm. Its scattering cross-section and scattering diagram are well-known to be rather spectrally sensitive in the vicinity of the magnetic resonance~\cite{KrasnokOE, Lukyanchuk13}. In particular, at some wavelengths, where magnetic and electric dipoles induced in the nanoparticle are almost equal and oscillate in phase, the silicon nanoparticle works as a Huygens source with suppressed backward scattering~\cite{KrasnokOE,Lukyanchuk13}. On the other hand, the scattering diagram can be tuned to the regime of suppressed forward scattering ("reverse" Huygens source), when magnetic and electric dipoles oscillate with the phase difference of $\pi$/2~\cite{KrasnokOE,Lukyanchuk13}.

\begin{figure}[!t]
\includegraphics[width=0.8\textwidth]{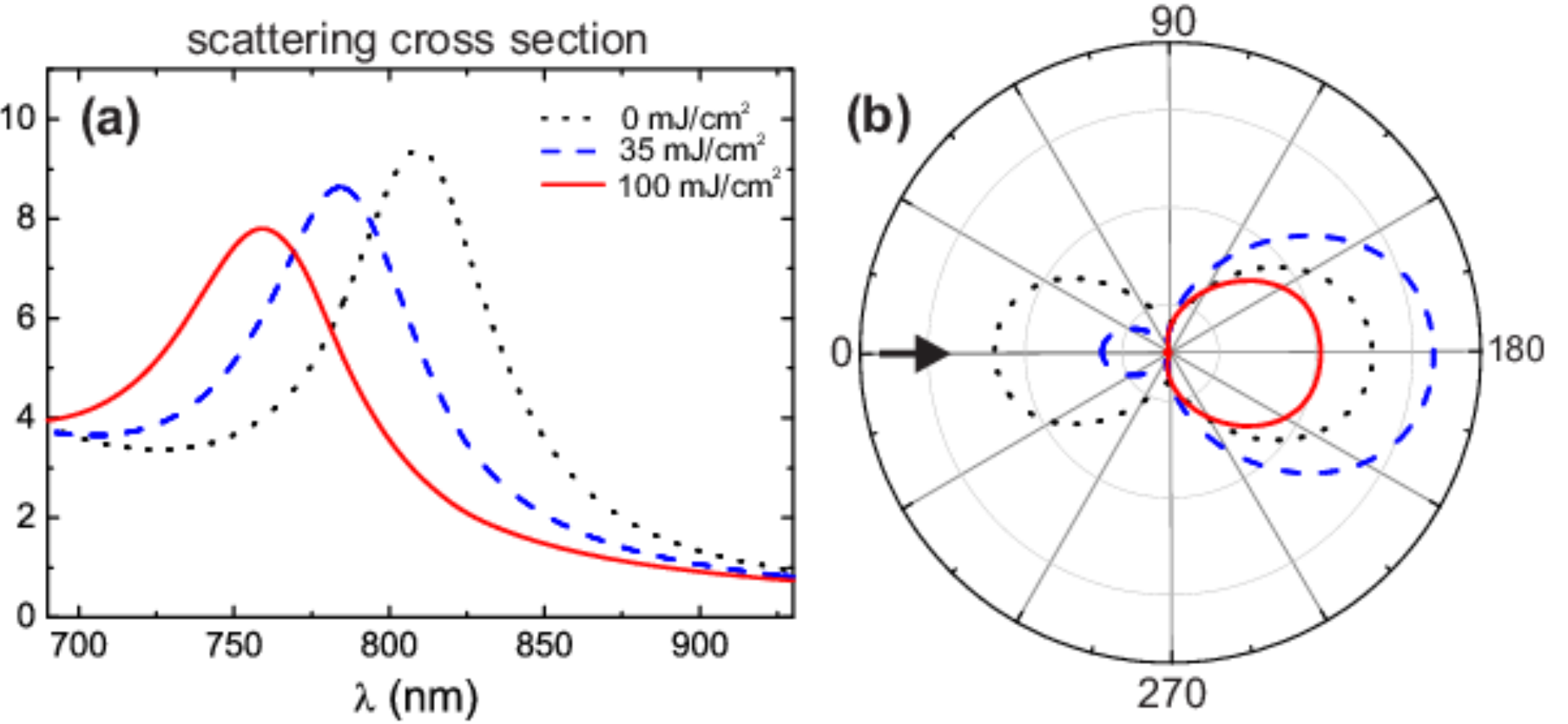}
\caption{Numerically calculated normalized scattering spectra (\textbf{a}) and scattering diagram at $\lambda$ = 800~nm (\textbf{b}) of the 210-nm silicon particle irradiated at $F_{\rm eff}$ = 0 (black curve), 35 mJ/cm$^2$ (red curve), and 100 mJ/cm$^2$ (blue curve). The dashed line corresponds to the wavelength of the pumping femtosecond laser.}\label{Fig4}
\end{figure}

The results of numerical simulations of scattering and scattering properties of the silicon nanoparticle in the vicinity of the magnetic dipole resonance are shown in Fig.~\ref{Fig3}. The calculated distributions of electric and magnetic fields indicate an emerging magnetic dipole moment in this range (Fig.~\ref{Fig3}a), while electric dipole moment is weak in the vicinity of 800 nm. At the points "b" and "d" the internal field represents the fields superposition of electric and magnetic nonresonant dipole modes. Since both scattering spectra (Fig.~\ref{Fig3}a) and scattering diagram (Fig.~\ref{Fig3}b-d) are strongly wavelength-dependent, it is possible to tune these two parameters by varying its dielectric permittivity.

Significant tuning of optical properties in the silicon nanoparticle near its magnetic dipole resonance requires time for the internal electromagnetic mode formation shorter than the corresponding electronic damping (thermalization) time and laser pulse duration. For the 210-nm silicon nanoparticle, the magnetic resonance mode has approximately the 50-nm full-width at the half- maximum, and, therefore, the \textit{Q}-factor is about 16, corresponding to the mode formation time in such open resonator of about 7 fs. Specifically, for dense EHP with ($\rho_{\rm eh}>10^{21}$ cm$^{-3}$ thermalization of free carriers proceeds over the characteristic plasma density-dependent times $\tau _{e} \sim$ 10--100 fs (see Eq. (3)) during the 100-fs pumping laser pulse. Hence, ultrafast plasma density-dependent changes in optical dielectric permittivity of the particle continuously occur during its laser pumping and are almost adiabatically followed by the internal electromagnetic mode.

To simulate numerically changes in optical properties of the photoexcited silicon sphere with the diameter \textit{D} = 210~nm at the wavelength $\lambda$ = 800~nm, we use the derived above dependencies of $\Delta{\rm Re}(\varepsilon)$ and $\Delta{\rm Im}(\varepsilon)$ on absorbed laser fluence $F_{\rm eff}$ (Fig.~\ref{Fig2}b). The considered range of absorbed fluences $F_{\rm eff}$ $<$ 100 mJ/cm$^2$ corresponds to non-destructive regime of the laser-particle interaction, but laser fluences are still high enough to generate rather dense EHP ($\rho_{\rm eh}$$\approx$1$\times$10$^{21}$ cm$^{-3}$) for efficient switching of the nanoparticle optical properties. As was mentioned above, the scattering diagram of the 210-nm nanoparticle is almost symmetric at $\lambda=800$~nm and $F_{\rm eff}\approx0$ (Fig.~\ref{Fig3}c), while the scattering cross section ($\sigma_{\rm ext}$) normalized on $\pi r^2$ has rather high value of about 9 (see Figs.~\ref{Fig3}a and~\ref{Fig4}a). The latter parameter is changed almost by three times with fluence increasing up to $F_{\rm eff}$ = 100 mJ/cm$^2$ at the fixed wavelength of 800~nm, owing to the strong shift of the peak position of the scattering spectrum (Fig.~\ref{Fig4}a). Its scattering diagram appears to be also very sensitive to the corresponding changes of the dielectric permittivity. At $\lambda\approx800$~nm, the transition from the typical dipole scattering diagram to the Huygens source forward scattering is observed in the fluence range $F_{\rm eff}=0-100$ mJ/cm$^2$ (Fig.~\ref{Fig4}b).

In order to prove the concept of manipulation by the electric and magnetic responses of a nanoparticle via optical photo-injection, we performed experiments with silicon nanoparticles of different sizes. First, we have fabricated arrays of truncated conical silicon nanoparticles with period $\Lambda \approx 800$ nm from an silicon film with thickness \textit{h}=220 nm deposited by PECVD on a fused silica substrate by means of inductive coupled plasma etching through electron-beam lithography prepared metal mask. The base ($r_{b}$) and top ($r_{t}$) radii of the nanoparticles were designed basing on preliminary numerical modeling, predicting the spectral position of the magnetic dipole resonance. Two different arrays were studied, composed by "near-resonance" ($r_{b} \approx$ 120 nm and $r_{t} \approx$ 70 nm) and "off-resonance" ($r_{b} \approx$ 70 nm and $r_{t} \approx$ 20 nm) nanoparticles. It should be noted here that the condition of Huygens source has been observed previously in visible and IR ranges for spherical, cylindrical and even conical semiconductor (silicon, GaAs, etc.) nanoparticles~\cite{staude2013tailoring, Novotny13, moitra2015large}.

\begin{figure}[!t]
\includegraphics[width=0.42\textwidth]{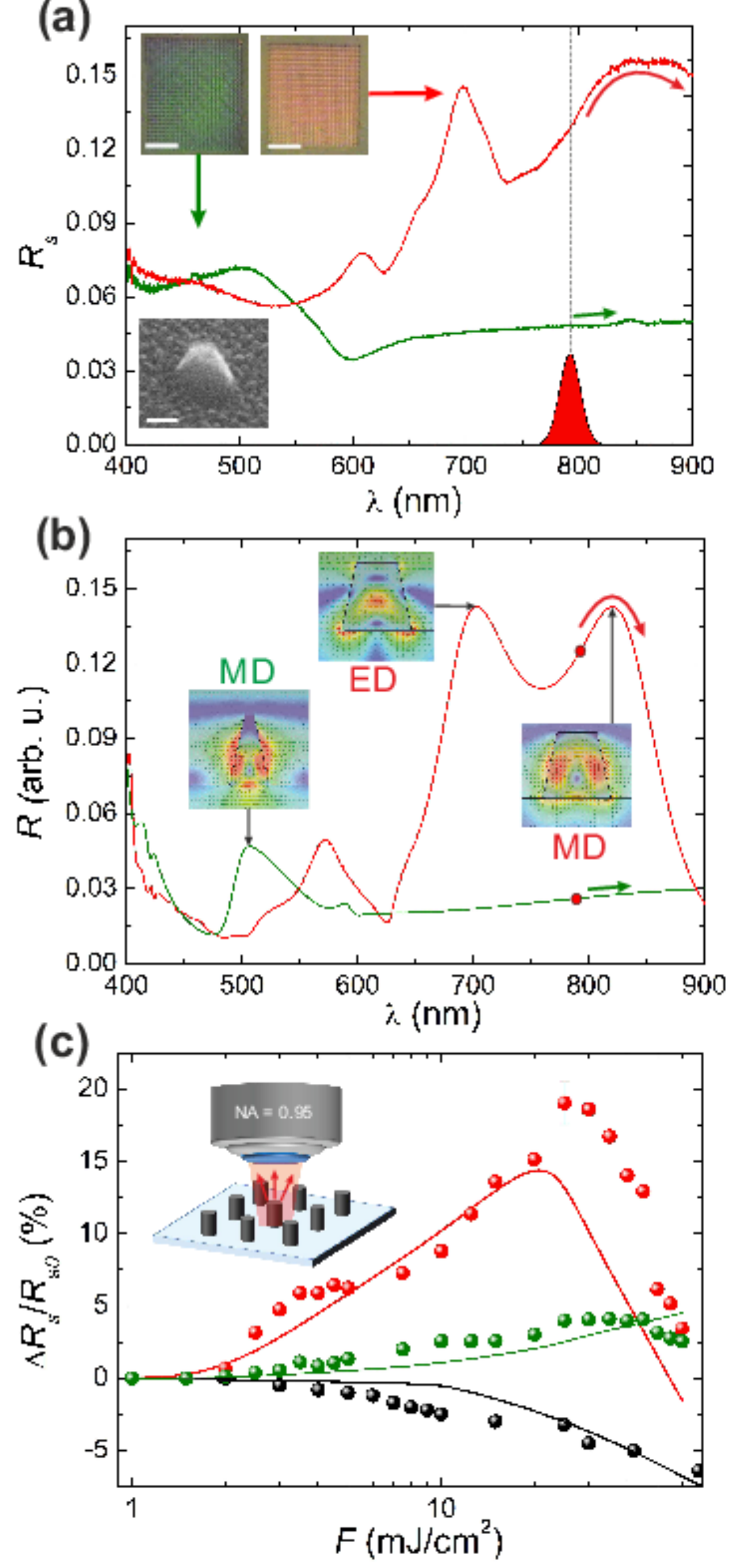}
\caption{Experimental (\textbf{a}) and numerical (\textbf{b}) reflection spectra of a truncated conical silicon nanoparticle (see the bottom left inset) with radii $r_{b} \approx$ 70 nm and $r_{t} \approx$ 20 nm (green curve), or $r_{b} \approx$ 120 nm and $r_{t} \approx$ 70 nm (red curve) with \textit{h} = 220 nm; spectrum of the femtosecond laser (right axis). The upper left insets in (\textbf{a}): optical image of silicon nanoparticles arrays. The bottom left inset in (\textbf{a}): SEM image of the "near-resonance" nanoparticle (the scale bar is 50 nm). The red filled curve in (\textbf{a}) represents the experimental spectrum of the femtosecond laser. The insets in (\textbf{b}) represent calculated electric field distributions near and inside the nanoparticles with the abbreviations: MD - magnetic dipole resonance, ED - electric dipole resonance. (\textbf{c}) Experimental (dots) and theoretical (solid lines) dependencies of normalized reflectance change on laser fulence for 220-nm thick silicon film (black), the "near-resonance nanoparticle" (red) and "off-resonance nanoparticle" (green). The values of \textit{R}$_{s0}$ are different for each sample and correspond to \textit{F} = 1 mJ/cm$^{2}$. The upper left inset: a schematic illustration of reflectance measurements from a single nanoparticle}\label{Fig5}
\end{figure}

The broadband spectral measurements of the reflected signal from a single nanoparticle ($R_{s}$) were carried out by means of strong focusing and collection of light ($\lambda$=400--900 nm) to a spectrometer (Horiba LabRam HR) through an achromatic objective with a numerical aperture NA=0.95. This objective allows to irradiate and collect light from an area of diameter $\sim 1.22\lambda/NA \leq \Lambda$. The resulting reflection spectra from different nanoparticles exhibit pronounced maxima (resonances) in different spectral ranges: one resonance is near $\lambda$ = 500 nm for the near-resonance nanoparticle and two resonances at $\lambda$ = 700 nm and $\lambda$ = 850 nm for the off-resonance nanoparticle (Fig.~\ref{Fig5}a). Our numerical modeling of reflection from silicon nanoparticles on SiO$_{2}$ substrate with the given dimensions and tabulated spectral dispersions~\cite{palik} reveals, that the observed maxima for near-resonance nanoparticle correspond to electric (700 nm) and magnetic (810 nm) dipole resonances, while the maximum in off-resonance nanoparticle spectrum has the magnetic response origin (Fig.~\ref{Fig5}b).

The nonlinear measurements of fluence dependent changes in the reflection signal from a single nanoparticle were carried out by using the objective (NA=0.95), similarly to the linear measurements. A commercial femtosecond laser system (TiF-100F, Avesta Poject) was used as an intense light source, providing 100-fs laser pulses at 790 nm central wavelength (FWHM $\approx$ 20 nm), with maximum pulse energy of 5 nJ and pulse duration of 100 fs at the repetition rate of 80 MHz. Laser energy was varied and controlled by an acousto-optical modulator (R23080-3-LTD, Gooch and Housego) and a power meter (FielfMax II, Coherent), respectively. The chosen energy range corresponds to fluences less than 100 mJ/cm$^{2}$, which is close but well below the melting threshold (\textit{F}$ \approx$ 250 mJ/cm$^{2}$~\cite{ionin2013Si}) and the observed damage fluence (\textit{F} $\approx$ 100 mJ/cm$^{2}$) of the nanoparticles. The reflected femtosecond laser pulses were collected by the focusing objective and directed to a calibrated CCD camera to measure reflected average fluence with high spatial resolution.

The wavelength of the femtosecond laser irradiation is located at the blue shoulder of the magnetic dipole resonance of the near-resonance nanoparticle, while it is far away from all resonances of off-resonance nanoparticle (Fig.~\ref{Fig5}a,b). The comparison of relative variation of reflectivity from the surface with the near-resonance nanoparticle, the non-resonant nanoparticle, and bulk silicon on laser fluence reveals the strongest response from the near-resonance nanoparticle. Specifically, the reflectance of the near-resonance nanoparticle exhibits 20$\%$
relative growth with incident fluence increasing up to $\approx$ 30 mJ/cm$^{2}$, while the reflection signal from the 220-nm thick silicon film goes down and the off-resonance nanoparticle reflection demonstrates intermediate behavior (Fig.~\ref{Fig5}b).

Such changes in reflection dependencies on laser fluence are totally governed by photo-injection of EHP, because we found good agreement of experimental dependencies of $\Delta$\textit{R$_{s}$}/\textit{R$_{s0}$} on incident fluence with the numerical simulations of changes in reflection signal (Fig.~\ref{Fig5}c), taking into account the calculated dielectric function dependence on laser fluence (Fig.~\ref{Fig2}). The growth of the reflectivity of the near-resonance nanoparticle is caused by the blue shift of the magnetic dipole resonance owing to the generation of EHP, decreasing dielectric permittivity of silicon with increase of laser fluence (Fig.~\ref{Fig2}).

In summary, photo-injection of dense electron-hole plasma in a dielectric nanoparticle, supporting a magnetic dipole resonance in the optical range, paves the way for effective light manipulation on subwavelength scale not only by tuning of scattering cross section of the nanoparticle, but also by tuning of its scattering diagram. In the frame of this concept, the 20$\%$
switching of reflection from a silica surface with a silicon nanoparticle photoexcited by a femtosecond laser pulse has been shown, enabling high-efficient light manipulation on the subwavelength scale.
In general, since the transient electronic dynamics under VIS-IR femtosecond laser photoexcitation is governed by similar physical processes (EHP generation, bands filling, bandgap renormalization, and  ion potential screening) in different semiconductors, the electron-hole plasma induced tuning of magnetic resonance can be applicable for broad range of materials and wavelengths. Therefore, tuning of dielectric permittivity of a single dielectric nanoparticle via photoexcitation of dense electron-hole plasma opens a novel class of ultracompact nanodevices with potentially ultrafast timescale, based on the diversity of "magnetic light" effects.

\begin{acknowledgement}
This work was financially supported by Russian Science Foundation (Grant 15-19-30023). The authors are thankful to A. Poddubny, M. Shcherbakov, and Yu. Kivshar for useful discussions. The authors also thank A.S. Gudovskikh for a-Si:H thin layer deposition by PECVD and I.A. Morozov for ICP etching of deposited a-Si:H thin layer.

The authors declare no competing financial interest.
\end{acknowledgement}


\providecommand{\latin}[1]{#1}
\providecommand*\mcitethebibliography{\thebibliography}
\csname @ifundefined\endcsname{endmcitethebibliography}
  {\let\endmcitethebibliography\endthebibliography}{}

\end{document}